\documentclass[prx,longbibliography,twocolumn,superscriptaddress,amsmath,amssymb]{revtex4-2}
\usepackage{amsmath,amssymb,amsfonts,mathtools,graphicx}
\usepackage[colorlinks=true,citecolor=blue,urlcolor=blue]{hyperref}
\graphicspath{{Figure/}}

\normalfont

\begin{document}

 \title{Sagnac-Loop-Reflector Fabry--P\'{e}rot Lattices for Modular 1D Topological Photonics}
	
\author{Siwoo Kim}
\thanks{These authors contributed equally to this work.}
\affiliation{Department of Physics, Korea Advanced Institute of Science and Technology, Daejeon 34141, Republic of Korea}

\author{Yung Kim}
\thanks{These authors contributed equally to this work.}
\affiliation{Department of Physics, Korea Advanced Institute of Science and Technology, Daejeon 34141, Republic of Korea}
\affiliation{Department of Mechanical Engineering, Korea Advanced Institute of Science and Technology, Daejeon 34141, Republic of Korea}

\author{Semin Choi}
\affiliation{Department of Physics, Korea Advanced Institute of Science and Technology, Daejeon 34141, Republic of Korea}

\author{Taeyeon Kim}
\affiliation{Department of Robotics and Mechatronics Engineering, Daegu Gyeongbuk Institute of Science and Technology, Daegu, Republic of Korea}

\author{Seungmin Lee}
\affiliation{Department of Robotics and Mechatronics Engineering, Daegu Gyeongbuk Institute of Science and Technology, Daegu, Republic of Korea}

\author{Kyoungsik Yu}
\affiliation{Department of Electrical Engineering, Korea Advanced Institute of Science and Technology, Daejeon 34141, Republic of Korea}

\author{Sangyoon Han}
\thanks{Corresponding author: s.han@dgist.ac.kr}
\affiliation{Department of Robotics and Mechatronics Engineering, Daegu Gyeongbuk Institute of Science and Technology, Daegu, Republic of Korea}

\author{Bumki Min}
\thanks{Corresponding author: bmin@kaist.ac.kr}
\affiliation{Department of Physics, Korea Advanced Institute of Science and Technology, Daejeon 34141, Republic of Korea}

\begin{abstract}

We introduce a modular silicon-photonic Fabry--P\'erot (FP) resonator lattice based on cascaded tunable Sagnac loop reflectors (SLRs). Each SLR is controlled by a single directional-coupler cross-coupling coefficient, enabling modular control of the effective lattice hoppings. As a representative example, alternating two SLR types maps the lattice onto the Su--Schrieffer--Heeger (SSH) model in the weak-coupling limit. We derive the Bloch dispersion via a transfer-matrix formulation and obtain an effective tight-binding Hamiltonian in the weak-coupling limit. $S$-parameter simulations of a 20-site lattice show an isolated midgap resonance with edge-localized power profiles in the topological phase, and disorder tests show robustness against symmetry-preserving hopping perturbations. Our results establish SLR-based FP lattices as a complementary platform for on-chip topological photonics.

\end{abstract}
 
\maketitle

\section{Introduction}

Coupled-resonator lattices realized in silicon photonics provide a versatile platform for band-structure engineering and on-chip emulation of tight-binding Hamiltonians~\cite{RevModPhys.91.015006,bogaerts2020programmable,shekhar2024roadmapping,pelucchi2022potential}. Microring-resonator arrays, in particular, have become a canonical implementation because of their compact footprint, mature design infrastructure, and natural compatibility with nearest-neighbor evanescent coupling, making them central to many studies of topological photonics and integrated photonic lattice Hamiltonians~\cite{leykam2020topological,lu2022chip,hafezi2011,hafezi2013imaging,zhao2019non,hotte2025integrated}. Scaling such arrays to larger site counts while maintaining spectrally resolved lattice features, however, remains practically challenging. Fabrication-induced variations in waveguide width and etch depth can introduce site-dependent resonance detunings, leading to inhomogeneous broadening and unintended shifts of the targeted spectral signatures~\cite{su2020silicon,shen2011electric}. In addition, when intracell and intercell couplings are set primarily by fixed directional-coupler geometries, achieving deterministic control over the effective hopping amplitudes generally requires iterative redesign and careful post-fabrication calibration of individual coupling sections~\cite{li2022extraction,jayatilleka2015wavelength}.

In this work, we introduce a one-dimensional Fabry-–P\'{e}rot (FP) resonator lattice based on cascaded tunable Sagnac loop reflectors (SLRs)~\cite{arianfard2023sagnac,wu2018advanced} as a complementary silicon-photonic platform for 1D topological photonics. In this architecture, each SLR acts as a compact reflective element whose response is primarily controlled by the cross-coupling coefficient of a single directional coupler, providing a local and modular handle on the effective hopping amplitudes. This contrasts with many ring-resonator lattices used for tight-binding emulation, where the coupling between neighboring site rings is typically mediated by a link ring and thus determined by two coupling sections. In the SLR-based FP architecture, by contrast, the effective hopping is controlled by a single coupler element. Because these coupler elements can, in principle, be independently characterized and tuned, for example through MEMS-based actuation~\cite{park2024fully}, the platform offers a potentially direct route to configuring lattice couplings across the device. In addition to this direct control of lattice couplings, the standing-wave FP implementation reduces the required resonator waveguide length to approximately half that of a traveling-wave ring resonator with the same resonant condition~\cite{sun2013tunable}. By alternating two SLR types within a unit cell, the structure implements a dimerized FP chain that reduces to the Su–-Schrieffer-–Heeger (SSH) model~\cite{su1979solitons,asboth2016short} in the weak-coupling limit, with the intracell and intercell couplings controlled by the two coupler settings.

We establish this correspondence using both transfer-matrix and weak-coupling descriptions. The transfer-matrix formulation~\cite{poon2004matrix} yields the Bloch dispersion and identifies the band-gap closing point associated with the topological phase transition when the two effective couplings become equal. In the weak-coupling limit, the round-trip dynamics reduce to an effective tight-binding Hamiltonian of SSH form. $S$-parameter simulations of a 20-site lattice, together with temporal coupled-mode-theory calculations~\cite{suh2004temporal}, reveal an isolated midgap resonance and edge-localized power profiles consistent with a topological edge state in the nontrivial regime. We further examine symmetry-preserving hopping disorder and find that the edge response remains localized for the perturbation strengths considered here.

\begin{figure*}[!t]
  \centering
  \includegraphics[width=\textwidth]{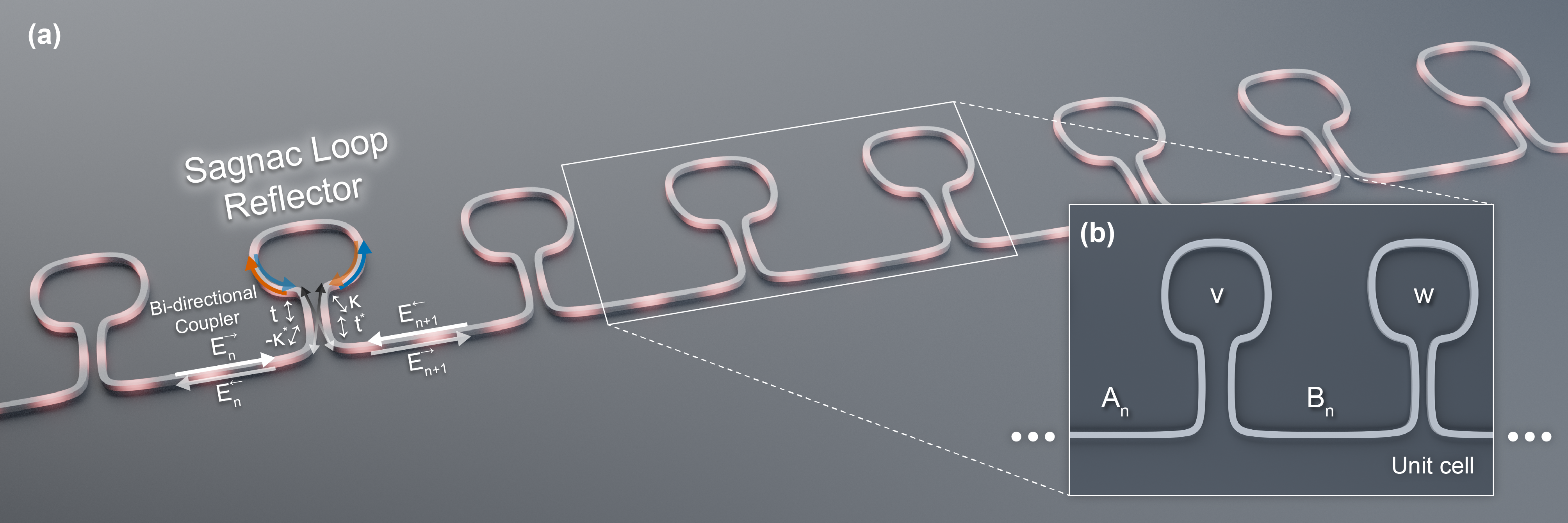}
  \caption{(a) Schematic of the FP lattice formed by cascading Sagnac loop reflectors, with waveguide segments between adjacent reflectors. One representative SLR is highlighted together with the forward and backward field amplitudes $E_n^{\rightarrow}$ and $E_n^{\leftarrow}$ and the directional-coupler coefficients $(t,\,t^{*})$ and $(\kappa,\,-\kappa^{*})$. (b) Dimerized unit cell composed of two SLRs of type $v$ and $w$, which define the effective intracell and intercell couplings of the lattice.}
  \label{fig:slr-fp_schematic}
\end{figure*}

\section{Results}

A FP lattice can be constructed by cascading SLRs, as illustrated in Fig.~\ref{fig:slr-fp_schematic}(a). In this architecture, each SLR acts as an effective reflective element, while two alternating SLR types define the dimerized unit cell shown in Fig.~\ref{fig:slr-fp_schematic}(b). For the present analysis, it is sufficient to describe an individual SLR by effective transmission and reflection amplitudes determined by the directional-coupler coefficients, up to a common propagation phase factor $e^{i\beta L}$, where $\beta$ is the propagation constant of the loop waveguide and $L$ is the SLR loop length, distinct from the inter-SLR segment lengths $L_A$ and $L_B$. We further assume that the $v$- and $w$-type SLRs share the same loop length $L$ and differ only in their coupler settings. Adopting the standard reciprocal coupler convention, the corresponding scattering relation is given by
\begin{equation}
    \begin{aligned}
        \begin{bmatrix}
            E_{n+1}^{\rightarrow} \\[2pt]
            E_{n}^{\leftarrow}
        \end{bmatrix}
        &=
        \mathrm{e}^{i\beta L}
        \begin{bmatrix}
            t_{\mathrm{SLR}} & r_{\mathrm{SLR}} \\
            - r_{\mathrm{SLR}}^{*} & t_{\mathrm{SLR}}^{*}
        \end{bmatrix}
        \begin{bmatrix}
            E_{n}^{\rightarrow} \\[2pt]
            E_{n+1}^{\leftarrow}
        \end{bmatrix},
        \\
        t_{\mathrm{SLR}} &\coloneqq |t|^{2}-|\kappa|^{2},
        \qquad
        r_{\mathrm{SLR}} \coloneqq 2 t^{*}\kappa ,
    \end{aligned}
    \label{eq:sagnac_matrix}
\end{equation}
where $t$ and $\kappa$ are the self- and cross-coupling coefficients of the directional coupler, respectively. In the lossless case ($|t|^{2}+|\kappa|^{2}=1$), the SLR amplitudes also satisfy the unitarity condition $|t_{\mathrm{SLR}}|^{2}+|r_{\mathrm{SLR}}|^{2}=1$. The subscript $n$ labels the scattering plane (or waveguide segment) between adjacent SLRs, while the superscripts $\rightarrow$ and $\leftarrow$ denote forward- and backward-propagating fields (see Appendix~A for the derivation). Moreover, the transfer-matrix description used hereafter is valid away from the singular point $t_{\mathrm{SLR}}=0$, which corresponds to exact 50:50 coupling.

Equation~\eqref{eq:sagnac_matrix} shows that the SLR acts as a compact and tunable reflective element whose effective transmission and reflection amplitudes are primarily controlled by a single coupler parameter, apart from the common propagation phase factor. In particular, the effective transmission amplitude $t_{\mathrm{SLR}}$ vanishes for a 50:50 coupler, $|t|=|\kappa|=1/\sqrt{2}$, while $|t_{\mathrm{SLR}}|\rightarrow 1$ and $|r_{\mathrm{SLR}}|\rightarrow 0$ in the limit $\kappa\rightarrow 0$. This local control of the SLR response provides the basis for implementing the alternating couplings required for the SSH-type FP lattice. As shown in Fig.~\ref{fig:slr-fp_schematic}(b), we take the unit cell to consist of a waveguide segment $A_n$, an SLR of type $v$, a second waveguide segment $B_n$, and an SLR of type $w$. This dimerized structure provides a natural setting for comparison with the SSH model, in which the intracell and intercell couplings alternate. To describe the field evolution from the left boundary of the $A_n$ segment to that of the $A_{n+1}$ segment, we use a transfer-matrix formulation \cite{poon2004matrix} in the basis
\[
\mathbf{E}_{n,A}\coloneqq
\begin{bmatrix}
E_{n,A}^{\rightarrow}\\
E_{n,A}^{\leftarrow}
\end{bmatrix}.
\]

\noindent The propagation across one unit cell is then written as
\begin{align}
    \mathbf{E}_{n+1,A}
    &= \mathbf{T}_{w}\,\mathbf{D}_{B}\,\mathbf{T}_{v}\,\mathbf{D}_{A}\,\mathbf{E}_{n,A},
    \notag\\
    \mathbf{T}_{\alpha}
    &=
    \frac{1}{t_{\mathrm{SLR},\alpha}}
    \begin{bmatrix}
        \mathrm{e}^{i\beta L} & r_{\mathrm{SLR},\alpha}\\
        r_{\mathrm{SLR},\alpha}^{*} & \mathrm{e}^{-i\beta L}
    \end{bmatrix},
    \qquad \alpha\in\{v,w\},
    \notag\\
    \mathbf{D}_{\mu}
    &=
    \begin{bmatrix}
        \mathrm{e}^{i\beta L_{\mu}} & 0\\
        0 & \mathrm{e}^{-i\beta L_{\mu}}
    \end{bmatrix},
    \qquad \mu\in\{A,B\},
    \label{eq:unitcell_tm}
\end{align}
where $\mathbf{T}_{\alpha}$ denotes the transfer matrix of the Sagnac loop reflector of type $\alpha$, with coefficients defined in Eq.~\eqref{eq:sagnac_matrix}, and $\mathbf{D}_{\mu}$ describes propagation through the waveguide segment associated with sublattice $\mu$. We assume a common propagation constant $\beta$ for the loop and inter-SLR waveguide sections. Throughout this work, we neglect both propagation loss and coupler loss, so that $\beta$ is real and $|t_{\mathrm{SLR},\alpha}|^{2}+|r_{\mathrm{SLR},\alpha}|^{2}=1$. As noted above, the transfer-matrix description is valid away from the singular point $t_{\mathrm{SLR},\alpha}=0$, corresponding to exact 50:50 coupling.

\begin{figure*}[!t]
  \centering
  \includegraphics[width=0.9\textwidth]{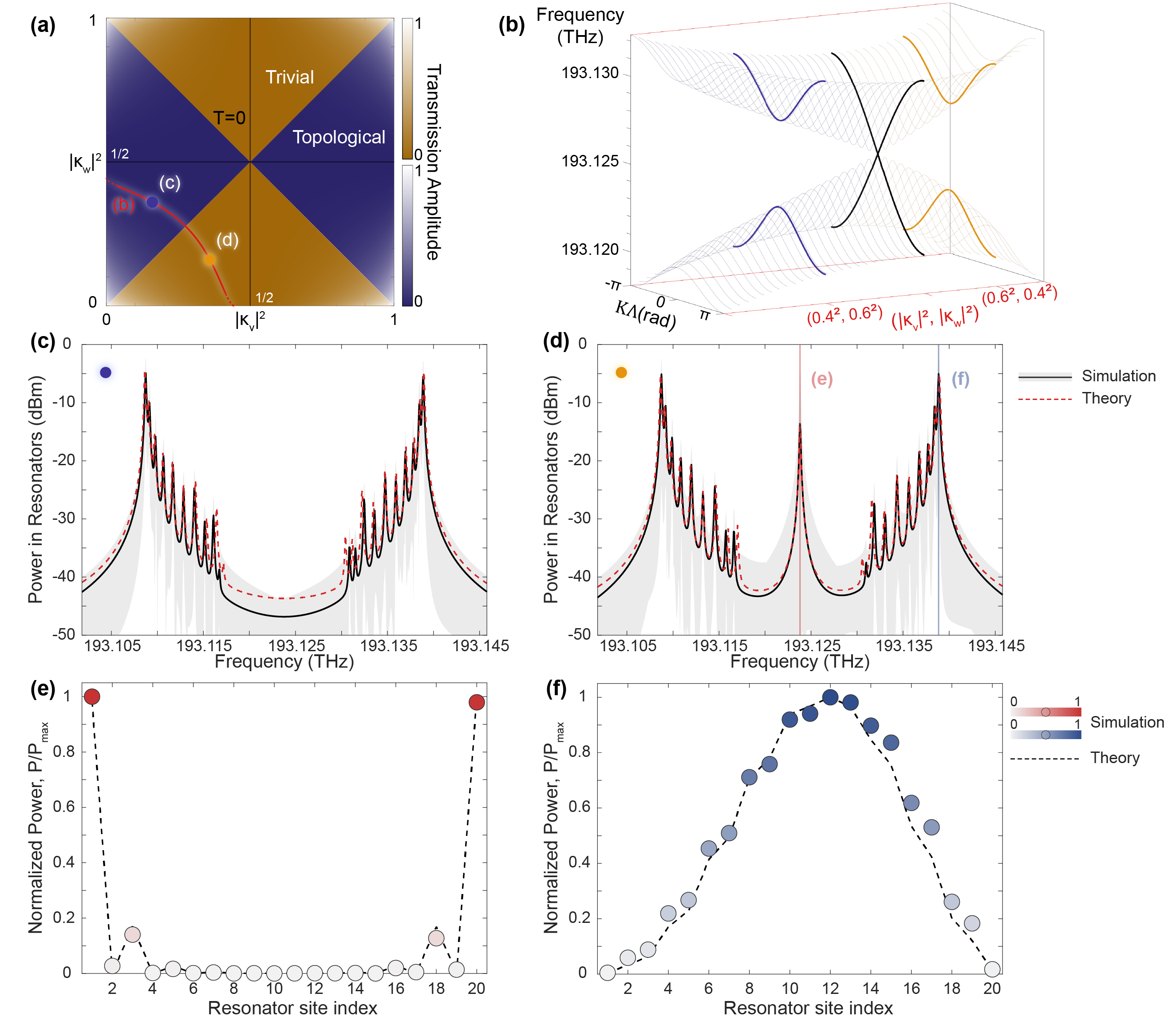}
  \caption{
    (a) Unit-cell transmission amplitude as a function of $(|\kappa_v|^2,\,|\kappa_w|^2)$, with red contours indicating the constant-transmission trajectory and the markers denoting the representative points $(|\kappa_v|,\,|\kappa_w|)=(0.4,\,0.6)$ and $(0.6,\,0.4)$.
    (b) Dispersion relation plotted as frequency versus Bloch phase $K\Lambda$ obtained by scanning $(|\kappa_v|,\,|\kappa_w|)$ along the solid red contour in (a).
    (c-f) Circuit-level simulation results (solid) and temporal coupled-mode-theory predictions (dashed) for a 20-site SLR--FP lattice. In (c) and (d), the solid curves show the site-averaged resonator power spectra, while the gray bands indicate the corresponding site-to-site range, bounded by the minimum and maximum simulated site powers at each frequency. (c) Trivial and (d) topological coupling configurations at the two representative points in (a). (e) Edge-mode and (f) bulk-mode site-resolved power profiles at the frequencies marked in (d).}
  \label{fig:total_result}
\end{figure*}

The unit-cell transfer matrix also determines the field transmission amplitude of a single unit cell. Figure~\ref{fig:total_result}(a) shows the resulting transmission amplitude as a function of $|\kappa_v|^2$ and $|\kappa_w|^2$. The two marked points, $(|\kappa_v|,\,|\kappa_w|)=(0.4,\,0.6)$ and $(0.6,\,0.4)$, are used below as representative trivial and topological configurations, respectively. The red contour indicates parameter pairs with the same unit-cell transmission amplitude, allowing the coupling ratio to be varied while keeping the unit-cell transmission fixed. For an infinite periodic lattice, Bloch's theorem gives $\mathbf{E}_{n+1,A}=\mathrm{e}^{iK\Lambda}\mathbf{E}_{n,A}$, where $K$ is the Bloch wave vector and $\Lambda$ is the lattice constant. In the lossless case, the unit-cell transfer matrix is unimodular, so its Bloch eigenvalues are $\mathrm{e}^{\pm iK\Lambda}$. The dispersion relation therefore follows from
\begin{equation}
    \cos(K\Lambda) = \frac{1}{2}
    \operatorname{Tr}\!\left(
    \mathbf{T}_{w}\mathbf{D}_{B}
    \mathbf{T}_{v}\mathbf{D}_{A}
    \right).
    \label{eq:dispersion_tmf}
\end{equation}

The corresponding dispersion is shown in Fig.~\ref{fig:total_result}(b), where the coupling condition is scanned along the solid red contour in Fig.~\ref{fig:total_result}(a). As $|\kappa_v|$ and $|\kappa_w|$ are varied across the line $|\kappa_v|=|\kappa_w|$, the band gap closes and then reopens. This behavior is consistent with the transition between the two dimerization patterns of the SSH model. Within this correspondence, the two SLR types set the effective intracell and intercell couplings, so that the regimes $|\kappa_v|<|\kappa_w|$ and $|\kappa_v|>|\kappa_w|$ correspond to the trivial and topological phases, respectively, for the unit-cell convention adopted here.

To connect the transfer-matrix description to an effective lattice model, we next consider the round-trip dynamics in the weak-coupling regime. Starting from the SLR scattering relation, the field amplitudes on sublattices $A$ and $B$ obey discrete-time round-trip equations in which coupling within a unit cell is mediated by the SLR of type $v$, whereas coupling between neighboring unit cells is mediated by the SLR of type $w$. When the effective SLR transmission amplitudes are small, the field envelopes vary slowly over a single round-trip time $\tau_{\mathrm{rt}}$, so the discrete update can be approximated by a continuous-time coupled-mode equation. In this limit, the amplitudes satisfy an effective nearest-neighbor lattice dynamics of the form
\begin{equation}
    \begin{aligned}
        \dot{E}_{n,A} &= -\epsilon\,E_{n,A} + J_v^* E_{n,B} + J_w E_{n-1,B},\\
        \dot{E}_{n,B} &= -\epsilon\,E_{n,B} + J_v E_{n,A} + J_w^* E_{n+1,A},
    \end{aligned}
    \label{eq:effective_cm}
\end{equation}
where
\begin{equation}
    J_v \coloneqq \frac{t_{\mathrm{SLR},v}}{\tau_{\mathrm{rt}}},
    \qquad
    J_w \coloneqq \frac{t_{\mathrm{SLR},w}}{\tau_{\mathrm{rt}}},
\end{equation}
and $\epsilon$ denotes a residual onsite-like term set by the round-trip phase and reflection coefficients. Under the reciprocal-coupler convention adopted here, $J_v$ and $J_w$ are real; we retain the complex-conjugate notation to facilitate comparison with the gauge-equivalent Hermitian form derived in Appendix~B. Equation~\eqref{eq:effective_cm} has the structure of a dimerized tight-binding chain, with $J_v$ and $J_w$ playing the roles of the intracell and intercell couplings. Under an appropriate phase-matching condition (see Appendix~B), the diagonal term $\epsilon$ becomes higher order in the weak-coupling parameter and can be neglected to leading order. The resulting dynamics are therefore SSH-type to leading order, and a gauge-equivalent Hermitian SSH representation is derived in Appendix~B. The SLR--FP lattice thus provides a weak-coupling realization of an SSH-type model in which the two alternating SLR types set the two hopping amplitudes. For Fig.~\ref{fig:total_result}(b) and all subsequent results, we set $L_A=L_B=L=800~\mu\mathrm{m}$ and $n_{\mathrm{eff}}=2.8$, and use the corresponding $\tau_{\mathrm{rt}}\approx1/\mathrm{FSR}=(0.1~\mathrm{THz})^{-1}$ in the effective model.

We next examine the finite-lattice response of the SLR--FP system using Lumerical simulations based on an $S$-parameter circuit model. For the finite-chain simulations, the terminal SLRs at the chain boundary were set to the 50:50 condition, $\kappa_{term}=1/\sqrt{2}$, thereby defining the boundary condition. In a typical photonic-integrated-circuit layout, excitation from a single edge port can bias the response toward modes with large overlap near the driven boundary, thereby obscuring the intrinsic mode-resolved response of the finite lattice. To reduce this effect, we use a weakly coupled tapping scheme in which each FP segment is coupled to an external port, so that the lattice response can be probed without relying on a single-edge input. The simulated multiport structure is modeled using temporal coupled-mode theory (TCMT)~\cite{suh2004temporal}, based on the effective lattice dynamics introduced above and a uniform external coupling rate for all tap couplers. We consider two 20-site lattices corresponding to the representative parameter sets in Fig.~\ref{fig:total_result}(a), namely $(|\kappa_v|,\,|\kappa_w|)=(0.4,\,0.6)$ and $(0.6,\,0.4)$. A broadband excitation is applied through the tap ports with a fixed phase profile chosen to probe the lattice modes in a spatially distributed manner. 

The resulting site-averaged power spectra are shown in Fig.~\ref{fig:total_result}(c) and (d). The solid curves in Fig.~\ref{fig:total_result}(c) and (d) denote the circuit-level simulation results, while the dashed curves show the corresponding TCMT predictions. The TCMT model reproduces the main spectral features of the circuit-level simulations for both parameter sets. For $(|\kappa_v|,\,|\kappa_w|)=(0.4,\,0.6)$, the spectrum shows a set of resonances distributed across the band. For $(|\kappa_v|,\,|\kappa_w|)=(0.6,\,0.4)$, an isolated midgap resonance appears inside the gap between the two bulk bands, consistent with a topological edge-state response of the SSH-type lattice. To examine the associated spatial profiles, we next use single-frequency excitation at the two frequencies marked in Fig.~\ref{fig:total_result}(d): the midgap resonance at $193.1238~\mathrm{THz}$ and a bulk-band frequency at $193.1390~\mathrm{THz}$. The resulting site-resolved power profiles are shown in Fig.~\ref{fig:total_result}(e) and (f). At the midgap resonance, the response is concentrated near the two boundaries, whereas the bulk-band excitation produces a more extended profile across the lattice. These results are consistent with the expected distinction between edge-like and bulk-like modes in the dimerized lattice.

\begin{figure}
  \centering
  \includegraphics[width=0.9\columnwidth]{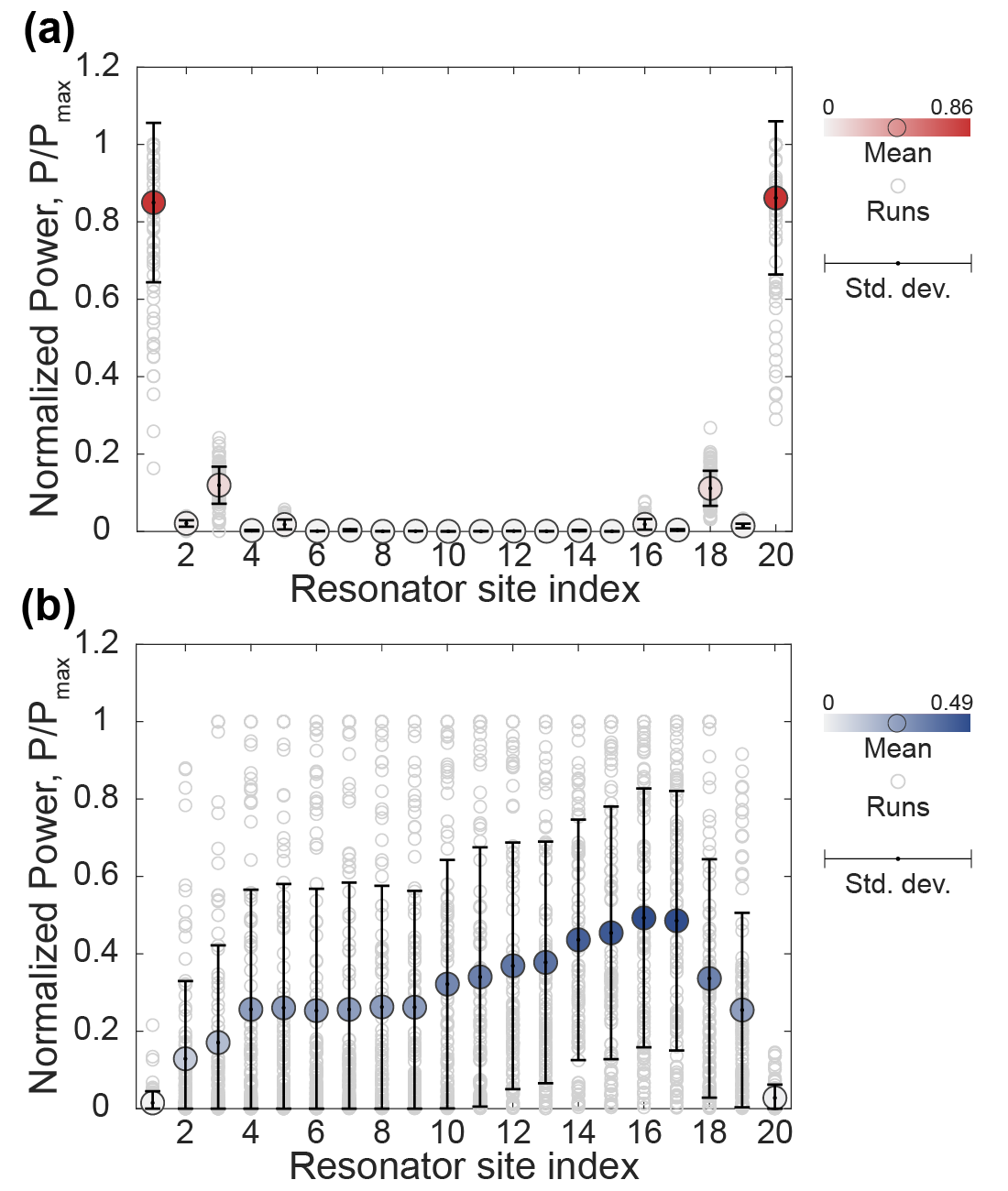}
  \caption{
  Disorder-averaged site-resolved power profiles for the topological coupling configuration in the presence of disorder applied to the raw coupler parameters: (a) excitation at the midgap frequency and (b) excitation at a bulk-band frequency. Filled markers denote the mean normalized site power over 100 disorder realizations, open circles denote individual realizations, and the error bars indicate $\pm 1\sigma$ across realizations, with the lower error bar clipped at zero because the normalized power is nonnegative. Independent Gaussian perturbations with standard deviation $\sigma=0.03$ are applied to the coupler cross-coupling coefficients $\kappa_{\alpha}$.
  }
  \label{fig:random_result_plot}
\end{figure}

We next examine the response of the SLR--FP lattice in the topological coupling configuration to disorder in the raw coupler parameters. Starting from the operating point $(|\kappa_v|,\,|\kappa_w|)=(0.6,\,0.4)$, we apply independent Gaussian perturbations with standard deviation $\sigma=0.03$ to the coupler cross-coupling coefficients $\kappa_{\alpha}$ and evaluate the site-resolved power profiles over 100 disorder realizations. This choice reflects the fact that the physical implementation is controlled directly through the raw coupler parameters rather than the effective SSH hopping amplitudes. In the lossless case, the effective couplings satisfy $J_{\alpha}\propto t_{\mathrm{SLR},\alpha}=1-2|\kappa_{\alpha}|^{2}$, so the regime $|\kappa_v|>|\kappa_w|$ corresponds to $|J_w|>|J_v|$, i.e. the topological side of the SSH-type correspondence adopted here. The disorder strength $\sigma=0.03$ was chosen so that the sampled coupler parameters remain predominantly within this regime. In practice, unequal inter-SLR separations can also introduce additional phase offsets. Here, however, we neglect such phase disorder and perturb only the coupler cross-coupling coefficients, thereby isolating the SSH edge-mode response to coupling disorder. Because the perturbations enter through the coupling sector rather than through onsite detunings, they act primarily as hopping disorder in the effective SSH description and preserve chiral symmetry to leading order. Within this setting, the midgap response is expected to be more robust against disorder than the bulk response.

Figure~\ref{fig:random_result_plot} shows the disorder-averaged site-resolved power profiles at the midgap frequency $193.1238~\mathrm{THz}$ and the bulk-band frequency $193.1390~\mathrm{THz}$, with error bars denoting $\pm 1\sigma$ across the 100 disorder realizations. At the midgap frequency (Fig.~\ref{fig:random_result_plot}(a)), the response remains concentrated near the two ends of the lattice, with only modest realization-to-realization variation. At the bulk-band frequency (Fig.~\ref{fig:random_result_plot}(b)), the spatial profile is distributed more broadly across the lattice and exhibits substantially larger variation across realizations. These results are consistent with the greater stability of the edge-like response against symmetry-preserving hopping disorder in the dimerized SLR--FP lattice~\cite{chen2020elementary}.

\section{Discussion}

We introduced a one-dimensional FP resonator lattice formed by cascading tunable Sagnac loop reflectors in silicon photonics and established its correspondence to an SSH-type lattice through transfer-matrix and weak-coupling analyses. Circuit-level $S$-parameter simulations and temporal coupled-mode theory consistently show an isolated midgap resonance and boundary-localized power profiles in the topological coupling configuration, supporting the validity of the effective lattice description in the parameter regime considered here.

Compared with evanescently coupled ring arrays, the SLR--FP lattice offers a more direct link between the effective couplings and individual directional couplers. In conventional ring arrays, the effective hopping strengths are determined by the geometry of the inter-ring coupling regions and are often inferred indirectly from the full lattice response \cite{vitullo2020coupling,li2022extraction}. In the SLR--FP lattice, by contrast, the relevant coupler cross-coupling ratios can in principle be characterized independently through standard transmission measurements. This per-element access is attractive for programmable implementations based on MEMS-actuated couplers \cite{park2024fully}, although a detailed experimental realization remains for future work.

The present analysis assumes lossless waveguides and identical segment lengths. In practice, propagation loss and path-length variations will introduce additional detunings and broadening. Still, our disorder analysis shows that disorder applied to the raw coupler parameters acts primarily as hopping disorder in the effective SSH description and preserves chiral symmetry to leading order, leaving the edge-like response substantially more stable than the bulk response, consistent with the robustness expected for SSH-type boundary modes under symmetry-preserving perturbations \cite{chen2020elementary}. Beyond the dimerized chain studied here, the same platform could be extended to longer-range SSH variants, Rice--Mele-type lattices, and non-Hermitian photonic lattices \cite{liu2022complex,RevModPhys.93.015005}. More broadly, the scalability of the architecture and its per-coupler programmability suggest that SLR--FP lattices provide a complementary route toward programmable topological photonics on silicon \cite{dai2024programmable,on2024programmable}.

\begin{acknowledgements}
This work is supported by the National Research Foundation of Korea (NRF) through the government of Korea (RS-2022-NR070636) and the Samsung Science and Technology Foundation (SSTF-BA240202). S.Y.H. acknowledges financial support from the National Research Foundation of Korea (NRF) grant funded by the Korea government (MSIT) (No. RS-2024-00352718).
\end{acknowledgements}

\appendix

\section{Derivation of the Sagnac Loop Reflector Scattering Matrix}

The Sagnac loop consists of a directional coupler and a closed waveguide loop, as shown in Fig.~\ref{fig:slr-fp_schematic}(a). For a directional coupler connecting two adjacent waveguides, the scattering matrix is
\begin{equation}
    \begin{bmatrix}
        E_{1}^{\mathrm{out}} \\[2pt]
        E_{2}^{\mathrm{out}}
    \end{bmatrix}
    =
    \begin{bmatrix}
        t & \kappa \\
        - \kappa^{*} & t^{*}
    \end{bmatrix}
    \begin{bmatrix}
        E_{1}^{\mathrm{in}} \\[2pt]
        E_{2}^{\mathrm{in}}
    \end{bmatrix},
    \label{eq:s_matrix_dc}
\end{equation}
where the subscripts label the two waveguides, and the superscripts denote the input and output ports. Under lossless and reciprocal conditions, the unitary coupler matrix in Eq.~\eqref{eq:s_matrix_dc} is fully specified by \(t\) and \(\kappa\), subject to \( |t|^{2}+|\kappa|^{2}=1 \).

Assuming identical phase accumulation for the two counter-propagating loop paths, propagation along the loop of length \(L\) adds a common phase factor to the two fields exiting the coupler. The loop propagation is therefore represented by \( \mathrm{e}^{i\beta L}\mathbf{I} \).

After a round trip in the loop, the two counter-propagating fields return to the coupler with their positions interchanged, which is represented by the permutation matrix \( \mathbf{P} \). Moreover, because the fields re-enter the coupler from the opposite side, the input and output ports are interchanged relative to the first pass. This geometric relabeling is captured by transposing the coupler matrix. The effective scattering matrix for the second pass is thus
\begin{equation}
    \mathbf{P}^{-1}
    \begin{bmatrix}
        t & \kappa \\
        -\kappa^{*} & t^{*}
    \end{bmatrix}^{\mathrm{T}}
    \mathbf{P}
    =
    \begin{bmatrix}
        t^{*} & \kappa \\
        -\kappa^{*} & t
    \end{bmatrix},
    \qquad
    \mathbf{P}=
    \begin{bmatrix}
        0 & 1\\
        1 & 0
    \end{bmatrix}.
\end{equation}
By cascading the first coupler, the loop propagation, and the transformed second-pass coupler, the overall scattering relation of the Sagnac loop is obtained as Eq.~\eqref{eq:sagnac_matrix}.

\section{Reduction of the SLR-–FP Lattice to an Effective SSH Model}

Starting from the discrete round-trip dynamics implied by Eqs.~\eqref{eq:sagnac_matrix} and \eqref{eq:unitcell_tm}, we consider the equal-length case $L_A=L_B\coloneqq L_{\mathrm{FP}}$ and write the field amplitudes on sublattices $A$ and $B$, after one discrete round-trip time $\tau_{\mathrm{rt}}$ as

\begin{widetext}
\begin{equation}
    \begin{aligned}
        E_{n,A}(t+\tau_{\mathrm{rt}})&\mathrm{e}^{-i\omega \tau_{\mathrm{rt}}}
        = -r_{\mathrm{SLR},w}\,r_{\mathrm{SLR},v}^{*}\,E_{n,A}(t)\mathrm{e}^{2i\beta (L+L_{\mathrm{FP}})}+ t_{\mathrm{SLR},v}^{*}\,E_{n,B}(t)\mathrm{e}^{i\beta (L+L_{\mathrm{FP}})} + t_{\mathrm{SLR},w}\,E_{n-1,B}(t)\mathrm{e}^{i\beta (L+L_{\mathrm{FP}})}\\
        E_{n,B}(t+\tau_{\mathrm{rt}})&\mathrm{e}^{-i\omega \tau_{\mathrm{rt}}}
        = -r_{\mathrm{SLR},v}\,r_{\mathrm{SLR},w}^{*}\,E_{n,B}(t)\mathrm{e}^{2i\beta (L+L_{\mathrm{FP}})}
        + t_{\mathrm{SLR},w}^{*}\,E_{n+1,A}(t)\mathrm{e}^{i\beta (L+L_{\mathrm{FP}})} + t_{\mathrm{SLR},v}\,E_{n,A}(t)\mathrm{e}^{i\beta (L+L_{\mathrm{FP}})}.
    \end{aligned}
    \label{eq:field_rt}
\end{equation}
\end{widetext}

In the weak-coupling regime, the amplitudes vary slowly over a single round trip. We therefore expand \( E_{n,\mu}(t+\tau_{\mathrm{rt}}) \) to first order in \( \tau_{\mathrm{rt}} \),
\begin{equation}
    \begin{aligned}
        E_{n,A}(t+\tau_{\mathrm{rt}}) \approx &
        E_{n,A}(t) + \tau_{\mathrm{rt}} \dot{E}_{n,A}(t)\\
        E_{n,B}(t+\tau_{\mathrm{rt}}) \approx &
        E_{n,B}(t) + \tau_{\mathrm{rt}} \dot{E}_{n,B}(t)
    \end{aligned}
    \label{eq:field_taylor}
\end{equation}
Substituting Eq.~\eqref{eq:field_taylor} into Eq.~\eqref{eq:field_rt}, we obtain the effective coupled-mode form \cite{steck2007quantum,liang2012optical}. The common round-trip phase factor \( \mathrm{e}^{i\beta (L+L_{\mathrm{FP}})} \) associated with \( t_{\mathrm{SLR},\alpha} \) and \( t_{\mathrm{SLR},\alpha}^{*} \) can be absorbed into the definition of the slowly varying amplitudes. This reproduces Eq.~\eqref{eq:effective_cm} in the main text, with
\begin{equation}
    \begin{aligned}
    \epsilon \coloneqq \frac{1}{\tau_{\mathrm{rt}}}\Bigl[r_{\mathrm{SLR},v}^{*}r_{\mathrm{SLR},w}e^{i\phi}+1\Bigr],
    \quad
    \phi \coloneqq 2\beta (L+L_{\mathrm{FP}})+\omega\tau_{\mathrm{rt}},
    \end{aligned}
    \nonumber
\end{equation}
\(J_v\) and \(J_w\) as defined in the main text.

In the weak-coupling regime, the Sagnac-loop transmission scales as \( t_{\mathrm{SLR},\alpha}=O(\delta) \) with \( \delta \ll 1 \), so that \( |J_v|, |J_w| = O(\delta) \). Unitarity then implies \( |r_{\mathrm{SLR},\alpha}| \simeq 1 - |t_{\mathrm{SLR},\alpha}|^{2}/2 \). Together with the phase-matching condition \(\phi + \mathrm{arg}(r{_\mathrm{SLR},w}/r{_\mathrm{SLR},v}) =(2n+1)\pi\), this gives \( |\epsilon| = O(\delta^{2}) \). The diagonal term \( \epsilon \) is therefore a higher-order correction and is neglected to leading order in the effective model. For ideally phase-matched couplers, this condition further reduces to \(\phi=(2n+1)\pi\), since \(\arg\!\bigl(r_{\mathrm{SLR},v}\bigr)=\arg\!\bigl(r_{\mathrm{SLR},w}\bigr)=\pi/2\).

For direct comparison with the SSH Hamiltonian, we redefine the effective couplings as \( v \coloneqq iJ_v \) and \( w \coloneqq iJ_w \), and identify the modal amplitudes \( a_{n,\mu} \) with \( E_{n,\mu} \) up to an overall normalization and phase convention. In the original convention, the corresponding evolution matrix is anti-Hermitian. However, by an appropriate rephasing of the reflection and transmission coefficients that preserves unitarity, the corresponding amplitudes satisfy
\begin{equation}
    \begin{aligned}
    \dot a_{n,A} = -i\bigl( w\,a_{n-1,B} + v^*\,a_{n,B}\bigr)\\
    \dot a_{n,B} = -i\bigl( v\,a_{n,A} + w^{*}a_{n+1,A} \bigr),
    \end{aligned}
\end{equation}
which can be written compactly as \( \dot{\mathbf a} = -i\,\mathbf H_{\mathrm{eff}}\mathbf a \), with Hermitian \( \mathbf H_{\mathrm{eff}} \). This equation is formally identical to the Heisenberg equation of motion generated by the SSH Hamiltonian,
\begin{equation}
    \hat H_{0} =
    \sum_{n} \left(
    v\,\hat{a}_{n,B}^{\dagger}\hat{a}_{n,A}
    +
    w\,\hat{a}_{n+1,A}^{\dagger}\hat{a}_{n,B}
    + \mathrm{h.c.}\right),
    \label{eq:ssh_hamiltonian}
\end{equation}
for which
\begin{equation}
    \dot{\hat a}_{n,\mu}
    =
    -\frac{i}{\hbar}\bigl[\hat a_{n,\mu}, \hat H_{0}\bigr].
    \label{eq:ssh_heisenberg}
\end{equation}
Setting \( \hbar = 1 \), the modal-amplitude equation and the operator equation take the same form. Here, \( \hat a_{n,\mu} \) and \( \hat a_{n,\mu}^{\dagger} \) denote the annihilation and creation operators on sublattice \( \mu \in \{A,B\} \) in the \( n \)th unit cell.

\bibliography{reference}

\end{document}